\documentclass{article}

\usepackage{amsmath}
\usepackage{graphicx}

\usepackage{oljour}

\begin{document}

% Bitte entfernen Sie das Kommentarzeichen vor dem Kürzel der
% Zeitschrift, in der Ihr Beitrag erscheinen soll.
% Please remove the comment char in the line that contains
% the shortcut for the journal where your submission shall
% appear

\journalname{it}  % (Information Technology)

% Englischer Titel
% Title in english
\title[en]{Computational Molecular Engineering as an emerging technology in process engineering}

% Titel des Beitrags (auf deutsch)
% Title of the submission (in german)
\title[de]{Computational Molecular Engineering als aufstrebende Technologie in der Verfahrenstechnik}

% Detaillierte Angaben zu allen Autoren des Beitrags:
% Bitte verwenden Sie für jeden Autor eine separate
% (begin/end) author-Umgebung.
% Detailed information about all authors of the submission:
% Please use for every author a separate
% (begin/end) author environment
\begin{author}
  \anumber{1}   %fortlaufende Nummer (running number)
  \atitle{}    %akademischer Titel (academic title)
  \firstname{Martin}  %Vorname(n) (first name(s))
  \surname{Horsch}    %Nachname (surname)
  \vita{$^\star$1982 (Stuttgart); 2006 Dipl.-Inf.\ (U.\ Stuttgart); 2010 Dr.-Ing.\ (U.\ Paderborn); 2011 Junior Professor for Computational Molecular Engineering (TU Kaiserslautern)}      %Kurzlebenslauf (short curriculum vitae)
  \institute{Laboratory of Engineering Thermodynamics (LTD), TU Kaiserslautern}   %Institutsname (name of institute)
  \street{Erwin-Schr\"odinger-Str.\ 44}    %Straße (street)
  \number{}    %Hausnummer (number)
  \zip{D-67663}      %Postleitzahl (zip code)
  \town{Kaiserslautern}
  \country{Germany}   %Land (country)
  \tel{+49 631 205 4028}      %Telefon (+COUNTRY-AREA-NUMBER: +49-341-49122701)
  \fax{+49 631 205 3835}      %Fax
  \email{martin.horsch@mv.uni-kl.de}     %E-Mail-Adresse (email address)
\end{author}

\begin{author}
   \anumber{2}
   \atitle{}
   \firstname{Christoph}
   \surname{Niethammer}
   \vita{$^\star$1983 (Stuttgart); 2009 Dipl.-Phys.\ (U.\ Stuttgart)}
   \institute{High Performance Computing Centre Stuttgart (HLRS)}
   \street{Nobelstr.\ 19}
   \number{}
   \zip{D-70569}
   \town{Stuttgart}
   \country{Germany}
   \tel{+49 711 685 87203}
   \fax{+49 711 685 65832}
   \email{niethammer@hlrs.de}
\end{author}

\begin{author}
  \anumber{3}   %fortlaufende Nummer (running number)
  \atitle{}    %akademischer Titel (academic title)
  \firstname{Jadran}  %Vorname(n) (first name(s))
  \surname{Vrabec}    %Nachname (surname)
  \vita{$^\star$1967 (Basel); 1992 Dipl.-Ing.\ (U.\ Bochum); 1996 Dr.-Ing.\ (U.\ Bochum); 1997--1999 OMC GmbH (D\"usseldorf); 2007 Habilitation (U.\ Stuttgart); 2009 Professor for Thermodynamics and Energy Technology (U.\ Paderborn)}      %Kurzlebenslauf (short curriculum vitae)
  \institute{Thermodynamics and Energy Technology (ThEt), University of Paderborn}   %Institutsname (name of institute)
  \street{Warburger Str.\ 100}    %Straße (street)
  \number{}    %Hausnummer (number)
  \zip{D-33098}      %Postleitzahl (zip code)
  \town{Paderborn}
  \country{Germany}   %Land (country)
  \tel{+49 5251 602 421}      %Telefon (+COUNTRY-AREA-NUMBER: +49-341-49122701)
  \fax{+49 5251 603 522}      %Fax
  \email{jadran.vrabec@upb.de}     %E-Mail-Adresse (email address)
\end{author}

\begin{author}
  \anumber{4}   %fortlaufende Nummer (running number)
  \atitle{}    %akademischer Titel (academic title)
  \firstname{Hans}  %Vorname(n) (first name(s))
  \surname{Hasse}    %Nachname (surname)
  \vita{$^\star$1960 (Landau); 1984 Dipl.-Ing.\ (U.\ Karlsruhe, TH); 1990 Dr.-Ing.\ (TU Kaiserslautern); 1995 Habilitation (TU Kaiserslautern); 1995--1998 BASF AG (Ludwigshafen); 1998 Professor for Thermodynamics and Thermal Process Engineering (U.\ Stuttgart); 2008 Professor for Thermodynamics (TU Kaiserslautern)}      %Kurzlebenslauf (short curriculum vitae)
  \institute{Laboratory of Engineering Thermodynamics (LTD), TU Kaiserslautern}   %Institutsname (name of institute)
  \street{Erwin-Schr\"odinger-Str.\ 44}    %Straße (street)
  \number{}    %Hausnummer (number)
  \zip{D-67663}      %Postleitzahl (zip code)
  \town{Kaiserslautern}
  \country{Germany}   %Land (country)
  \tel{+49 631 205 3497}      %Telefon (+COUNTRY-AREA-NUMBER: +49-341-49122701)
  \fax{+49 631 205 3835}      %Fax
  \email{hans.hasse@mv.uni-kl.de}     %E-Mail-Adresse (email address)
\end{author}

% E-Mail-Adresse für die Korrespondenz des Verlags zu diesem
% Beitrag
% Corresponding address (for the publisher) for this submission
\corresponding{martin.horsch@mv.uni-kl.de}

% englische Zusammenfassung des Beitrags
% english summary of the submission
\abstract{The present level of development of molecular force field methods is
assessed from the point of view of simulation-based engineering,
outlining the immediate perspective for
further development and highlighting the newly emerging discipline
of \textit{Computational Molecular Engineering (CME)} which makes
basic research in soft matter physics fruitful for industrial applications.
Within the coming decade, major breakthroughs can be reached
if a research focus is placed on \textit{processes at interfaces},
combining aspects where an increase in the
accessible length and time scales due to massively parallel high-performance
computing will lead to particularly significant improvements.}

% % deutsche Zusammenfassung des Beitrags
% % german summary of the submission
\zusammenfassung{Der aktuelle Entwicklungsstand molekularer
Kraftfeldmethoden wird vom Standpunkt des simulationsgest\"utzten
Ingenieurwesens beurteilt, indem Perspektiven f\"ur die unmittelbare
Zukunft herausgearbeitet werden. Dabei ist insbesondere die neu
entstehende Disziplin des \textit{Computational Molecular Engineering (CME)}
zu beachten, die Ergebnisse aus der physikalischen
Grundlagenforschung f\"ur die industrielle Anwendung nutzbar macht.
Im kommenden Jahrzehnt sind hier gr\"o\ss{}ere Durchbr\"uche zu
erwarten, wenn in der Forschung ein Schwerpunkt auf \textit{Prozesse an
Grenzfl\"achen} gesetzt wird. Die Kombination aus den gr\"o\ss{}eren
L\"angen- und Zeitskalen, die durch massiv-paralleles
H\"ochstleistungsrechnen erschlossen werden, wird auf diesem Gebiet
zu besonders ma\ss{}geblichen Fortschritten f\"uhren.}

% bis zu sechs Keywords (englisch)
% up to six keywords (english)

% bis zu sechs Schlagwörter (deutsch)
% up to six keywords (german)
\schlagwort{Molekulardynamik, simulationsgest\"utztes Ingenieurwesen, Modellierung und Simulation}

% Widmung
% dedication
% \dedication{}

% Die nachfolgenden Angaben werden in der Regel vom Verlag eingetragen.
% The following information is usually supplied by the publisher.
%======================================================================
\received{21st May 13}
\accepted{}
\volume{}
\issue{}
\class{}
\Year{2013}
%======================================================================

\maketitle

% Hier bitte den Inhalt des Artikels einfügen.
% Please insert the content of your submission here.

\end{multicols}

\section{Introduction}

\begin{multicols}{3}

\noindent
\textit{Computational Molecular Engineering (CME)} is a novel discipline
within the broad spectrum of simulation-based en\-gineering and
high-performance computing, aiming at adapting molecular force field methods,
which were developed within the soft matter physics and
thermodynamics communities \cite{Allen1989}, to the needs of
industrial users in chemical and process engineering. 
Only in recent years have sufficently accurate molecular models
become available for a wide variety of fluids, and it is only now
that, due to massive parallelization, molecular simula\-tions of
complex nanoscopic systems (approaching the micro\-scale) are feasible
with a reasonable computational effort \cite{GHV12, SHUS10}.
The present article briefly comments on the present level of
development in CME (Section \ref{sec:cmetoday}) as well as the
massively parallel molecular simulation program \textit{$\ell$s1
mardyn} (Section \ref{sec:devel}).
In Section \ref{sec:cmetomorrow}, the conclusion is
drawn that an increase in the length scale accessible by molecular
simulation also requires an extension of the simula\-ted time scale to
facilitate a substantial progress.

\end{multicols}

\begin{figure}[t]
\centering
\includegraphics[width=12.5cm]{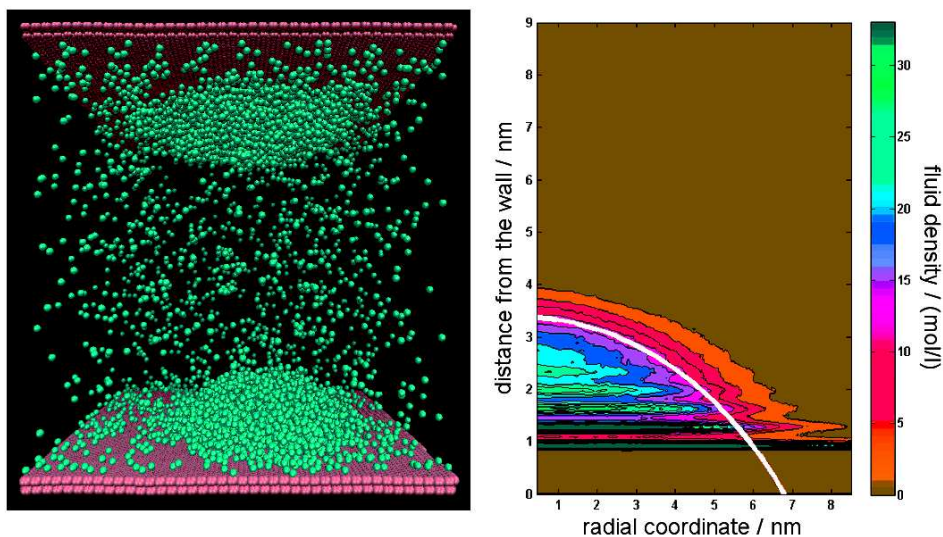}
\caption{MD simulation snapshot (left) and average fluid
density contour plot (right) for a sessile argon droplet on a
solid substrate. The simulation was conducted
with \textit{$\ell$s1 mardyn}.}
\label{fig:becker}
\end{figure}

\section{Computational Molecular Engineering today}
\label{sec:cmetoday}

\begin{multicols}{2}

\noindent
We witness today the progress from molecular
simulation as a theoretical and rather academic method to CME as an
arsenal of robust tools intended for practical use, which e.g.\
supplements or replaces experiments that are hazardous or hard to conduct
\cite{UNRAL07}. This follows the general pattern by
which engineering increasingly drives scientific development in areas
originating from pure chemistry and physics, building on
substantial basic research efforts, as soon as they have become ripe for
technical application.

The degree of sophistication of molcular force field methods and the
complexity of the simulated systems varies considerably
between the various fields of application. In particular, the
interdependence of elementary thermodynamic
properties such as pressure, density, temperature, enthalpy, and composition
can be reliably obtained by simula\-ting homogeneous systems that
contain up to 1 000 molecules \cite{Allen1989}. With
relatively little additional effort, higher-order derivatives of the
free energy (e.g.\ heat capacities or the speed of sound) are
accessible as well \cite{Lustig2011}; the case is similar for
mechanical properties of solid materials \cite{Roesch2009}.
By Grand Equilibrium or Gibbs ensemble
simulation, vapour-liquid equilibria
between homogeneous bulk phases, i.e.\ without an interface between
them, can be efficiently and accurately sampled \cite{DESLGGMBHV11, UNRAL07}.
Systems where a phase boundary is explicitly present
can also be treated, cf.\ Fig.\ \ref{fig:becker}. Such simulations require more
molecules, so that finite-size effects can be
isolated \cite{Binder97}, and longer computations (i.e., with more
simulation steps) need to be carried out, since fluid interfaces
often relax more slowly than the homogeneous bulk fluid and exhibit
more significant fluctuations, e.g.\ capillary waves, on a long time
scale.

This facilitates a modelling approach that has turned out to be
particularly fruitful in recent years: Thereby,
the electrostatic features of a molecular model, i.e.\ the choice
of parameters for point charges, dipoles or quadrupoles, are
determined from quantum chemical calculations.
United-atom sites interacting by the Lennard-Jones potential are
employed for intermolecular repulsion and dispersive London forces \cite{Allen1989},
also known as van der Waals forces. The corresponding potential
parameters are adjusted to optimize the overall agreement
with experimental data \cite{EVH08a}. These models, hence, though
simple, enable accounting for the most important types of molecular
interactions separately, including hydrogen bonding \cite{GHV12}.
Furthermore, they account for the structuring effects of the
interactions in the fluid (local concentrations, radial distribution
functions etc.) in a self-consistent way. This distinguishes them
from other approaches for describing fluid properties and explains
the fact that such models yield reliable extrapolations with respect to two
different aspects:
First, to conditions far beyond
those where the experimental data for the parameter fit were
determined; second, to a wide variety of fluid properties
which were not considered during parametrization at all \cite{EVH08a}.

Once pure component models are available, the extension to mixtures
is straightforward. Mixing rules are available for predicting the
unlike interaction parameters. If suitable experimental data are
available, adjustable binary parameters can be employed to improve
mixture models. This concept can also be applied to model\-ling
fluid-wall interactions, cf.\ Fig.\ \ref{fig:becker}. Furthermore, transferable pair potentials
are available which directly map functional groups to the model
parameters of corresponding single-atom or united-atom interaction
sites \cite{MS99}. In this way, molecular simulation can deploy its
predictive power, on the basis of a physically sound model\-ling
approach, even where the available set of experimental data reaches
its limits.
 
 % models are fit to VLE data

\end{multicols}

\begin{figure}[h]
\centering
\includegraphics[width=8.333cm]{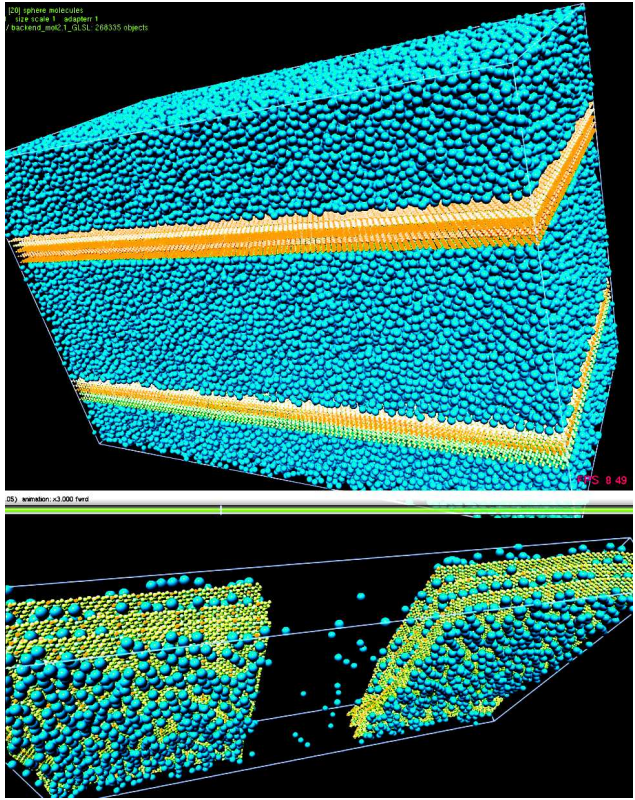}
\caption{Top: MD simulation snapshot for Couette shear flow of
methane in a graphite nanopore \cite{HVBH09}.
Bottom: Entrance effects, adsorption/desorption kinetics, and
permeability of fluid methane in nanoporous carbon, employing
non-equilibrium MD simulation \cite{HVBH09}. The simulations were conducted
with \textit{$\ell$s1 mardyn}.}
\label{fig:membranes}
\end{figure}

\begin{multicols}{3}

Both Monte Carlo (MC) and molecular dyna\-mics (MD) simulation are
suitable for determining most thermophysical properties:
MC simulation evaluates an ensemble average by stochastically
generating a representative set of configurations, i.e.\ position and
momentum coordinates of the molecules. Thereby, MC simulation uses
the Metropolis algorithm (which is randomized), whereas MD simulation
computes a trajectory segment by integrating Newton's equations
of motion (which are deterministic). If the same molecular force
field is used, temporal and ensemble averaging lead to consistent
results, since all thermodynamically relevant systems are at least
quasi-ergodic \cite{Allen1989}. MC simulation does not rely on time
and does not require an explicit computation of momentum coordinates,
which is advantageous for simula\-ting adsorption and
phase equilibria \cite{UNRAL07}; in these and similar cases, the most effective
methods involve grand-canonical or quasi-grand-canonical ensembles
with a varying number of molecules \cite{Binder97}, where MD simulation has the
disadvantage that momentum coordinates have to be determined for
molecules that are inserted into the system. For more complex
properties, however, e.g.\ regarding non-equilibrium states
and the associated relaxa\-tion processes,
time-dependent phenomena become essential, so that MD is the
preferred simula\-tion approach (cf.\ Fig.\ \ref{fig:membranes}).

Scientifically and technically, all preconditions for the introduction
of molecular simulation in an industrial environment are now
fulfilled \cite{GHV12}. Organizational aspects relevant for this
process include institutional support, the active
interest and involvement of both corporate and academic partners, and
channelling of the effort to a few simulation codes, at most, rather
than reinventing the HPC wheel again and again. In this respect, the
development in Great Britain can serve as a positive example, where
a community centered around the \textit{Computational Collaboration
Project 5} develops and applies the
\textit{DL\_POLY} program. An example for successful collaboration between
academia and industry can be found in the United States, where the
\textit{Industrial Fluid Properties Simulation Challenge} also
attracts international attention and participation \cite{EVH08a}.
However, the corresponding programming efforts are highly fragmented:
Parallel developments are attempted based on the \textit{Amber},
\textit{CHARMM}, \textit{LAMMPS}, \textit{NAMD}
and \textit{MCCCS Towhee} codes, among
many others \cite{MS99, PBWGTVCSKS05, Plimpton95}.

At present, however, the German CME community
arguably constitutes the best environment for mastering
the qualitative transition of molecular simulation from a scholarly
academic occupation to a key technology in industrial-scale fluid
process engineering. Its institutional structure guarantees an
orientation towards industrial use and successfully integrates
engineering with high performance computing. It is within this
framework that a consistent toolkit encom\-passing two major components
is developed: The \textit{ms2} program (i.e.\ \textit{\underline{m}olecular
\underline{s}imulation: \underline{second} generation}), intended for thermophysical
properties of bulk fluids -- the interested reader is referred to Deublein
et al.\ \cite{DESLGGMBHV11} -- and \textit{$\ell$s1 mardyn},
which is briefly introduced below.

\end{multicols}

\section{Large systems in molecular dynamics}
\label{sec:devel}

\begin{multicols}{2}

\noindent
The novel MD simulation code \textit{$\ell$s1 mardyn} (i.e.\
\textit{\underline{$\ell$}arge \underline{s}ystems \underline{first}: \underline{m}olecul\underline{ar} \underline{dyn}amics}) aims at expanding the
temporal and spatial range of scales accessible to molecular
simulation, with a focus on inhomogeneous systems (e.g.\ at
interfaces) and non-equilibrium thermodynamics. In
many re\-levant cases, such as those involving vapour liquid coexistence,
the molecule distribution may be very heterogeneous
and change over time in an unpredictable way (e.g.\ during phase decomposition or multi-phase
flow). The design of \textit{$\ell$s1 mardyn} was oriented towards three
major objectives: Modularity of the structure, interdisciplinar
collaboration within the development process, and transferability of
the code base to diverse and heterogeneous HPC architectures.

Molecular simulations with system dimensions that far exceed the
cut-off radius, beyond which a mean-field approach is employed for
the inter\-molecular interactions, are most efficiently parallelized by space
decomposition schemes. Thereby, the simulation volume is subdivided into
smaller subvolumes (one for each process) that ideally carry the same
load \cite{Bernard1999}. Finding an optimal balance
requires a method that estimates
the load corresponding to the possible decompositions on the fly,
since the particle density distribution can vary significantly over
simulation time.
In \textit{$\ell$s1 mardyn}, an interface class for
the domain decomposition scheme permits the generic implementation of
different load balancing strategies operating on spatial
subdomains. The linked-cell data structure, which
is used to find neighboring molecules, introduces a subdivision of
the simulation volume into small cells (with dimensions on the order
of the cut-off radius) that are also used as basic volume units for
the decomposition.

\begin{figure}[b]
\centering
\includegraphics[width=8.333cm]{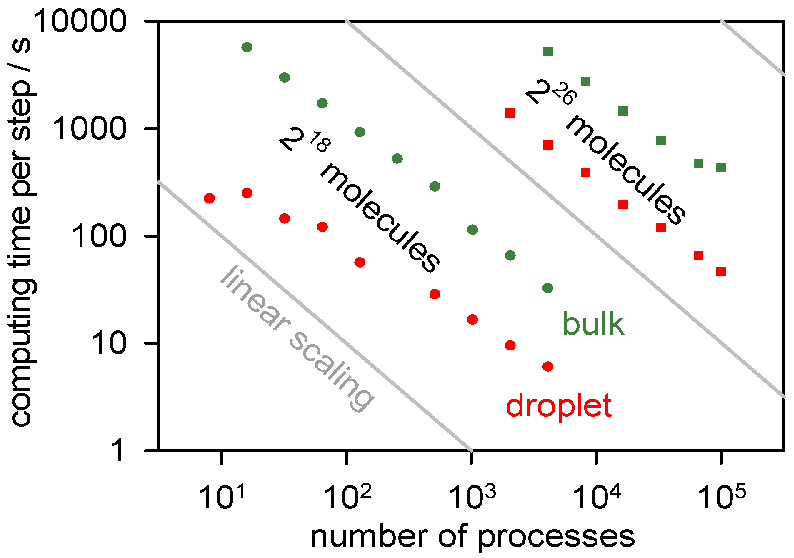}
\caption{Strong scaling of the \textit{$\ell$s1 mardyn} program on
the \textit{hermit} supercomputer for homogeneous (bulk) and
heterogeneous (droplet) scenarios.}
\label{fig:strong-scaling}
\end{figure}

On the basis of the computational cost for each of
the cells, load balancing algorithms can group cells together such
that $N_\textnormal{p}$ subvolumes with approximately equal load are created, where
$N_\textnormal{p}$ is the number of processing units.
A hierarchical tree-based approach (similar to $k$-dimensional
trees) turned out to be the most adequate dynamic load
balancing method; this concept was previously known to be suitable
for parallelizing particle-based simulations with short-range
interactions \cite{Bernard1999}. Thereby, the
simulation volume is recursively bisected by planes
with alternating orientation (normal to the $x$, $y$, $z$,
$x$, \dots\ axis). This process is repeated until each
process is assigned one cuboid subdomain.
With this technique, an excellent scalability was obtained even for
heterogeneous scenarios on up to $10^5$ processing units of the
\textit{hermit} supercomputer at the High Performance Computing
Centre Stuttgart (HLRS), cf.\ Fig.\ \ref{fig:strong-scaling}.

\end{multicols}

% \section{Impact of massively parallel high-performance computing}
\section{Conclusion}
\label{sec:cmetomorrow}

\begin{multicols}{3}

\noindent
From a computational point of view, large MC or MD simulations can
better be tackled than MD simulations of processes over a relatively
long time span. By far the largest part of the numerical effort is
required for evaluating the force field, a task which can be
efficiently distributed over multiple processes, as discussed above.
In contrast, the temporal evolution along a trajectory
through the phase space cannot be parallelized due to its inherently
sequential nature.
In the past, this has repeatedly led developers of highly performant
simulation codes to boast of the number of (low-density gas) molecules that
they succeeded in loading into memory as the single considered
benchmark criterion \cite{GK08}.

However, large and
\textit{interesting} systems generally also require more simulation
\textit{time}. Industrial users will hardly care
how many trillions of molecules can be
simulated for a dilute homogeneous gas over a few picoseconds, or even less.
From the point of view of thermodynamics and fluid
process engineering, the criterion for the world record in molecular
simulation should not be
the number of molecules $N$, but rather an exponent $\ell$ such that
e.g.\ within a single day, at least $N = 10^{3\ell}$ molecules in a condensed
state were simulated over at least $10^{\ell + 4}$ time steps. This would
promote a proportional increase of the accessible length and time
scales, which is what real-life applications require. 

By pushing this frontier forward, a wide spectrum of novel
scale-bridging simulation approaches will become feasible, paving the
way to a rigorous investigation of many size-dependent effects,
which on the microscale may be qualitatively different from the
nanoscale. Following this route, major breakthroughs will be reached
within the coming decade, assuming that a research focus is placed on
processes at interfaces.
By focussing on such applications,
cf.\ Fig.\ \ref{fig:membranes}, an increase in the accessible length and
time scale due to massively parallel high-performance computing will
lead to particularly significant improvements, opening up both
scientifically and technically highly interesting fields such as
microfluidics (including turbulent flow), coupled heat and mass
trasfer, and design of functional surfaces to an investigation on the
molecular level.
 
\end{multicols}
 
\begin{multicols}{2}

\noindent
\textit{Acknowledgment.} The present work was conducted under the
auspices of the Boltzmann-Zuse Society of Computational Molecular
Engineering (BZS). The authors would like to thank
M.\ F.\ Bernreuther, M.\ Buchholz, W.\ Eckhardt, S.\ V.\ Lishchuk and S.\ Werth for their contributions to developing the \textit{$\ell$s1 mardyn} MD simulation program,
C.\ Avenda\~n{}o, S.\ Becker, S.\ Eckelsbach, H.\ Frentrup,
P.\ Klein and E.\ A.\ M\"uller for helpful advice and discussions,
as well as DFG for funding SFB 926 (MICOS).

% \bibliographystyle{acm}
% \bibliography{it-bungartz} 

\end{document}